\documentstyle[aps,12pt,tighten]{revtex}
\begin{document}
\draft
%\preprint{}
\input{epsf}

\title{ The $\overline{p} p \rightarrow  \pi^0\pi^0 $ Puzzle}

\author{W. A. Perkins}
\address{(Perkins Advanced Computer Systems) Auburn, California }

\maketitle

\begin{abstract}
According to conventional theory, the 
annihilation reaction $\overline{p} p \rightarrow \pi^0 \pi^0 $ 
cannot occur from a $\overline{p} p$ atomic S state. 
However, this reaction occurs so readily for antiprotons stopping in 
liquid hydrogen, that it would require 30\% P-wave annihilations.
Experimental results from other capture and $\overline{p} p$ 
annihilation channels show that the fraction of P-wave 
annihilations is less than 6\% in agreement with theoretical 
expectations. An experimental test to determine whether this reaction 
can occur from an atomic S state is suggested. If indeed this reaction is
occurring from an atomic S state, then certain neutral 
vector mesons should exhibit a $\pi^0 \pi^0 $ decay mode, 
and this can also be tested experimentally. 
\end{abstract}


Because of parity conservation and symmetry of identical bosons with
respect to their interchange, the 
annihilation reaction $\overline{p} p \rightarrow \pi^0 \pi^0 $ 
cannot occur from a $\overline{p} p$ atomic S state. 
However, experiments dating back to the early 70's have shown 
that the reaction occurs readily for antiprotons stopping in 
liquid hydrogen. Although this mode is allowed from atomic P-state 
annihilations, the measurements with the Crystal Barrel
detector of the branching ratio for this reaction 
(when compared with
the $\overline{p} p \rightarrow \pi^+ \pi^-$ branching ratio) 
would require 30\% P-wave annihilations. Experimental results 
from $\pi^- p, K^- p,$ and $\Sigma^- p$ capture and other 
$\overline{p} p$ annihilation channels show that the 
fraction of P-wave 
annihilations is less than 6\% in agreement with theoretical 
expectations. Thus, the experimental 
evidence strongly indicates that 
the annihilation into $\pi^0\pi^0$ is occurring from an
atomic S state. We will discuss an experimental test involving
X-rays in coincidence with the reaction, which can 
determine whether or not this reaction is occurring predominantly 
from an atomic S state and other tests involving vector meson
decays. We will also consider how conventional theory
could be modified to allow this reaction to occur 
from an atomic S state.

The eigenvalues of parity and charge parity of a 
fermion-antifermion pair are given
by~\cite{roman},

\begin{eqnarray}
\omega_{Parity} = (-1)^{X + 1}, \nonumber \\ 
\omega_{Charge \; parity} = (-1)^{X + s}, 
\label{eqnpbarp}
\end{eqnarray}
where $X $ is the relative orbital angular momentum of the 
two particles and $s$ is the spin of 
the fermion-antifermion system. The eigenvalues of parity and 
charge parity of a 
two pion system are given by~\cite{roman},

\begin{eqnarray}
\omega_{Parity} = (-1)^Y, \nonumber \\ 
\omega_{Charge \; parity} = (-1)^{Y}, 
\label{eqnpipi}
\end{eqnarray}
where $Y$ is the relative orbital angular momentum 
of the two pions. 
For the $ \pi^0 \pi^0 $ system there are further constraints. 
Because of Bose statistics the state of two identical pions must 
be symmetric under interchange. Thus, $Y$ must be even and 
both parity 
and charge parity must be $+1$ for the $ \pi^0 \pi^0 $ system. 
Considering only low angular momentum states (i.e., $J = 0,1$),
we obtain Tables I-III for initial and final states. 

\begin{table}
\caption{ Initial state of $ \overline{p} p $  Atom }
\begin{tabular}{cccc}
State&$J$&Parity&Charge parity \\
\tableline
$^1S_0$&0&-1&+1 \\
$^3S_1$&1&-1&-1 \\
$^3P_0$&0&+1&+1 \\
$^1P_1$&1&+1&-1 \\
$^3P_1$&1&+1&+1 \\
\end{tabular}
\end{table}
Using conservation of parity, charge parity, and total angular 
momentum in matching the initial and final states, we obtain the 
allowed reactions:

\begin{eqnarray}
\overline p p \: (^3P_0) \rightarrow \pi^0 \pi^0 (S_0), \nonumber \\
\overline p p \: (^3S_1) \rightarrow \pi^+ \pi^- (P_1), \nonumber \\
\overline p p \: (^3P_0) \rightarrow \pi^+ \pi^- (S_0).
\label{eqnallow} 
\end{eqnarray}
Thus we see that the reaction $\overline{p} p \rightarrow  
\pi^0\pi^0 $ cannot occur from an atomic S state of the 
$\overline{p} p $ system.

\begin{table}
\caption{ Final state of $  \pi^0 \pi^0 $ System }
\begin{tabular}{ccccc}
State&$J$&Parity&Charge parity&Comment \\
\tableline
$S_0$&0&+1&+1& Y=0 \\
\end{tabular}
\end{table}

We will briefly review what happens when an antiproton or some 
other negatively charged particle is slowed down in liquid hydrogen 
and is captured in a Bohr orbit by a proton. This is 
illustrated in Fig.~\ref{f1}. 
Typically, the incoming negatively charged 
particle is initially captured in an orbit with principle 
quantum number 
$ n \approx 30 $ and with high orbital angular momentum, $X$. 
Collisional deexcitations and radiative transitions transform 
the atom to 
lower $n$ and  $X$ values. The electrically neutral atom
can then penetrate 
neighboring atoms and experience the electric field of the protons.
This causes Stark effect transitions between 
the degenerate orbital angular momentum states. The rates for 
radiative transition and nuclear absorption (or annihilation)
from P states are small in 
comparison with the rate that the Stark effect 
populates the S state.
Since S-state absorption (or annihilation) 
can happen from high $n$ values,
the atom is unlikely to deexcite to low $n$ values for which P state
nuclear absorption (or annihilation) is more important. 

\begin{table}
\caption{Final state of $  \pi^+ \pi^-$ System }
\begin{tabular}{ccccc}
State&$J$&Parity&Charge parity&Comment \\
\tableline
$S_0$&0&+1&+1& Y=0 \\
$P_1$&1&-1&-1&Y=1  \\
\end{tabular}
\end{table}

\epsfxsize=5.0in \epsfbox{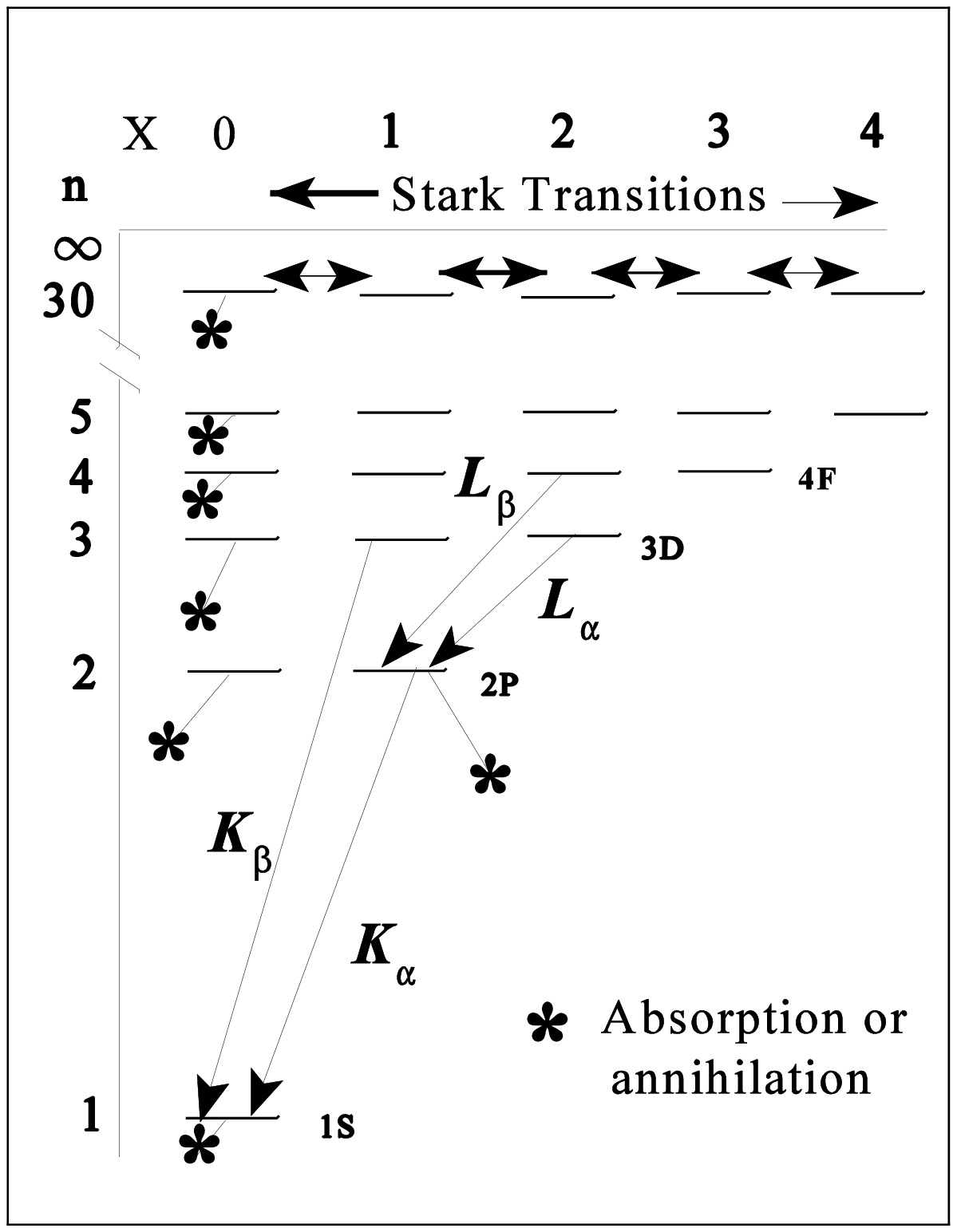}

\begin{figure}[t]
\caption{Levels of atomic orbital states for a negatively charged 
particle and a proton with 
principle quantum number $n$ and orbital angular momentum $X$. 
It shows the effect of Stark transitions on different $X$ states, 
radiative deexcitations, and levels
from which nuclear absorption or annihilation are likely.}
\label{f1}
\end{figure}

Thus, according to theory~\cite{day,leon-bethe} absorption will 
occur predominantly from S states for $\pi^-$ and $K^-$. In 1960, 
Desai~\cite{desai} concluded, 
``Rough calculations indicate that for protonium 
also the capture will take place predominantly from S states.''

There is strong experimental evidence that S-state capture dominates 
in liquid $H_2$. The reactions $ \pi^- p $~\cite{fields},  
$ K^- p $~\cite{cresti}, and 
$ \Sigma^- p $~\cite{burnstein} have been studied. Since these 
negatively-charged particles decay, one can determine the nuclear 
absorption time by observing the fraction
which decay. The cascade times are about
two orders of magnitude shorter than would be required
for radiative deexcitation. 
Because the antiproton does not decay, such a measurement is not 
possible. Since  
the short cascade times for $\pi^-, K^-,$ and $\Sigma^-$ cannot be 
explained without recourse to the Stark effect, the Stark effect 
must also play a role in the $\overline p p$ case.

There is some direct evidence of S-state domination in
 $\overline{p} p $ reactions. It has been determined~\cite{bizzarri1} 
that $\overline{p} p \rightarrow K K < 6\% \: $ P-wave with a 95\%
confidence level. 
From the $\rho $ decay angular distribution, 
it has been determined~\cite{foster,bizzarri2} that  
$\overline{p} p \rightarrow  \pi^+ \pi^- \pi^0 < 5 \% $
P-wave. Thus the experimental evidence strongly supports 
S-state domination for $\overline{p} p $ reactions. 

We will now look at the experimental results for 
$\overline{p} p \rightarrow  \pi \pi $. 
In liquid hydrogen the branching ratio for 
$\overline{p} p \rightarrow  \pi^+\pi^- $ 
is $ (31 \pm 1) \times 10^{-4}$~\cite{crystal2}, 
while measurements of the branching ratio for 
$\overline{p} p \rightarrow  \pi^0\pi^0 $ are given in 
Table IV. 

The experimental results obtained
with the Crystal Barrel detector 
are likely to be the most accurate. 
As they noted in their paper~\cite{crystal1}, 
``Owning to our large detection efficiency and small background 
our result is least likely influenced by undetected systematic 
errors. The reliability of the result is strengthened by the 
internal consistency of a large set of two-body branching ratios 
measured with the Crystal Barrel detector and their 
agreement with previous determinations, especially with bubble 
chamber data.''

One can calculate the fraction of P-wave annihilations for 
$ \overline{p} p \rightarrow \pi \pi $ as follows,

\begin{equation}
Fraction \; P \! \! - \! wave 
= {( \overline{p} p \rightarrow \pi^+ \pi^-)_P
+ (\overline{p} p \rightarrow \pi^0 \pi^0)_P  \over
( \overline{p} p \rightarrow \pi^+ \pi^-)_{S \& P}
+ (\overline{p} p \rightarrow \pi^0 \pi^0)_P }.
\label{eqnfrac}
\end{equation}
By assuming charge independence,
\begin{equation}
( \overline{p} p \rightarrow \pi^+ \pi^-)_P
= 2 \times (\overline{p} p \rightarrow \pi^0 \pi^0)_P,
\label{eqnchind}  
\end{equation}
we obtain the \% P wave given in Column 2 of Table IV. 
The 55\% P-wave, shown in Table IV for the Crystal Barrel collaboration,
is obviously too high, and they decided to use
another method~\cite{crystal1}, involving the P-wave fraction for 
$ \overline{p} p \rightarrow \pi^+ \pi^- $ 
determined from a measurement with gaseous 
hydrogen at NTP, giving 29\% P wave for their result.

\begin{table}
\caption{ Branching Ratio for
 $ \overline{p} p \rightarrow \pi^0 \pi^0 $  }
\begin{tabular}{cccc}
Measured value& \% P-wave&Year&Reference \\
\tableline
$(4.8 \pm 1.0) \times 10^{-4}$&$40\%$&1971&Devons {\it et. al.}
~\cite{devons} \\
$(1.4 \pm 0.3)\times 10^{-4}$&$13\%$&1979&Bassompierre {\it et. al.}
~\cite{bassom}\\
$(6 \pm 4) \times 10^{-4}$&$49\%$&1983&Backenstoss {\it et. al.}
~\cite{backen} \\
$(2.06 \pm 0.14) \times 10^{-4}$&$19\%$&1987&Adiels {\it et. al.}
~\cite{adiels} \\
$(2.5 \pm 0.3) \times 10^{-4}$&$22\%$&1988&Chiba {\it et. al.}
~\cite{chiba}\\
$(6.93 \pm 0.43) \times 10^{-4}$&$55\%$&1992&Crystal Barrel
~\cite{crystal1} \\
$(2.8 \pm 0.4) \times 10^{-4}$&$25\%$&1998&Obelix
~\cite{obelix} \\
\end{tabular}
\end{table}

Since even the 29\% P-wave fraction is much too high, Batty~\cite{batty}, 
trying to reduce this value, modified the earlier 
calculations of Reifenr\"{o}ther and E. Klempt~\cite{reinfen}
using the Borie and Leon model~\cite{borie}.  
He fitted the experimental data for $\overline{p} p \rightarrow \pi \pi$
and $\overline{p} p \rightarrow K K $ at several target 
densities. Without enhancement factors he obtained 27\% for the P-wave
fraction at liquid hydrogen density. 
With enhancement of annihilations
from fine structure states over that expected from
a statistical population and some adjustment of the parameters, 
he could fit the data (including the Crystal Barrel result) 
with a P-wave fraction of just 13\% at liquid hydrogen density. 
Even with this heroic effort, the 13\% P-wave fraction 
is still too high when compared with the 
upper limit of 6\% P-wave obtained from the results of 
other $\overline{p} p$ 
annihilations~\cite{bizzarri1,foster,bizzarri2}. 

There are two other experiments 
indicating anomalously large fractions of P-wave
($75\%$) involve antiproton annihilation at rest in liquid 
deuterium~\cite{gray,bridges}, although results from a third experiment 
were consistent with 
0\% P wave~\cite{angel}. Again the percentage P-wave
calculation is based on charge independence and the theoretical
argument that $\overline{p} d \rightarrow  n \pi^0\pi^0 $
cannot occur from an atomic S state.

How can one explain these anomalies? 
Since the results of experiments from all other capture and 
annihilation reactions point to S-state domination,
the most reasonable conclusion is that 
$\overline{p} p \rightarrow \pi^0 \pi^0 $ {\it is occurring from 
an atomic S state}. This would have been the conclusion
long ago if it did not violate conventional theory.

Although seven different groups have measured the branching ratio
for $\overline{p} p \rightarrow \pi^0 \pi^0 $, showing the importance
of this unexpectedly large branching ratio, no 
direct test has been performed to determine 
whether the reaction could be occurring from an
atomic S state. Such an experimental test can be made by setting 
up an initial $\overline{p} p $ atomic S state and looking for 
the $\pi^0 \pi^0$ final state. 
The method is illustrated in Fig.~\ref{f2}. 
As discussed earlier, in liquid $H_2$ the Stark effect causes 
transitions to S states at high n-values, where annihilation 
occurs more readily than deexcitation. One can 
decrease the effect of Stark transitions by using $H_2$ gas 
at NTP, and thereby observe the deexcitation radiation.

The coincidence of L and K X-rays from protonium, shows that 
the atom is in the 1S state. The energy of the K X-rays is 
between 9.4 KeV ($K_{\alpha}$) and 12.5 KeV ($K_{\infty}$), 
while energy of the L X-rays is between 1.7 KeV and 3.1 KeV. 
The energy of M X-rays is between 0.5 KeV and 1.3 KeV. Thus, 
the X-rays from the different transitions tend to be separated. 

The Asterix Collaboration has detected $K_{\alpha}$ X-rays in 
coincidence with L X-rays~\cite{asterix2}. The experiment 
we are proposing 
is very similar, but requires the triple coincidence of L and K 
X-rays from protonium and the $\pi^0 \pi^0$ annihilation mode. 
The detection of such events can prove that the annihilation 
reaction is occurring from an atomic S state.

As discussed above, even without this crucial experiment, 
there is significant evidence that the standard theory concerning 
this reaction may not be valid. 
We will now look at what could be wrong with the theory. 
To match the initial $^3S_1$ state of $\overline p p$, the
$\pi^0 \pi^0$ system needs a $P_1$ state with parity $= -1$ and 
charge parity $= -1$ 
as shown on the second line of Table III for the 
$\pi^+ \pi^-$ system. 
In conventional theory, the $\pi^0 \pi^0$ system has parity 
$= +1$ and 
charge parity $= +1$ for all states.

The big difference in those eigenvalues between the
$\pi^+ \pi^-$ system and the $\pi^0 \pi^0$ system in conventional
theory is caused by the two $\pi^0$'s being identical while the
$\pi^+$ and $\pi^-$ are not. Thus, it must be possible for one
$\pi^0$ to exist in one internal states while the other $\pi^0$ 
is in another internal state. This would result in
the two $\pi^0$'s not always being identical.
\epsfxsize=5.6in \epsfbox{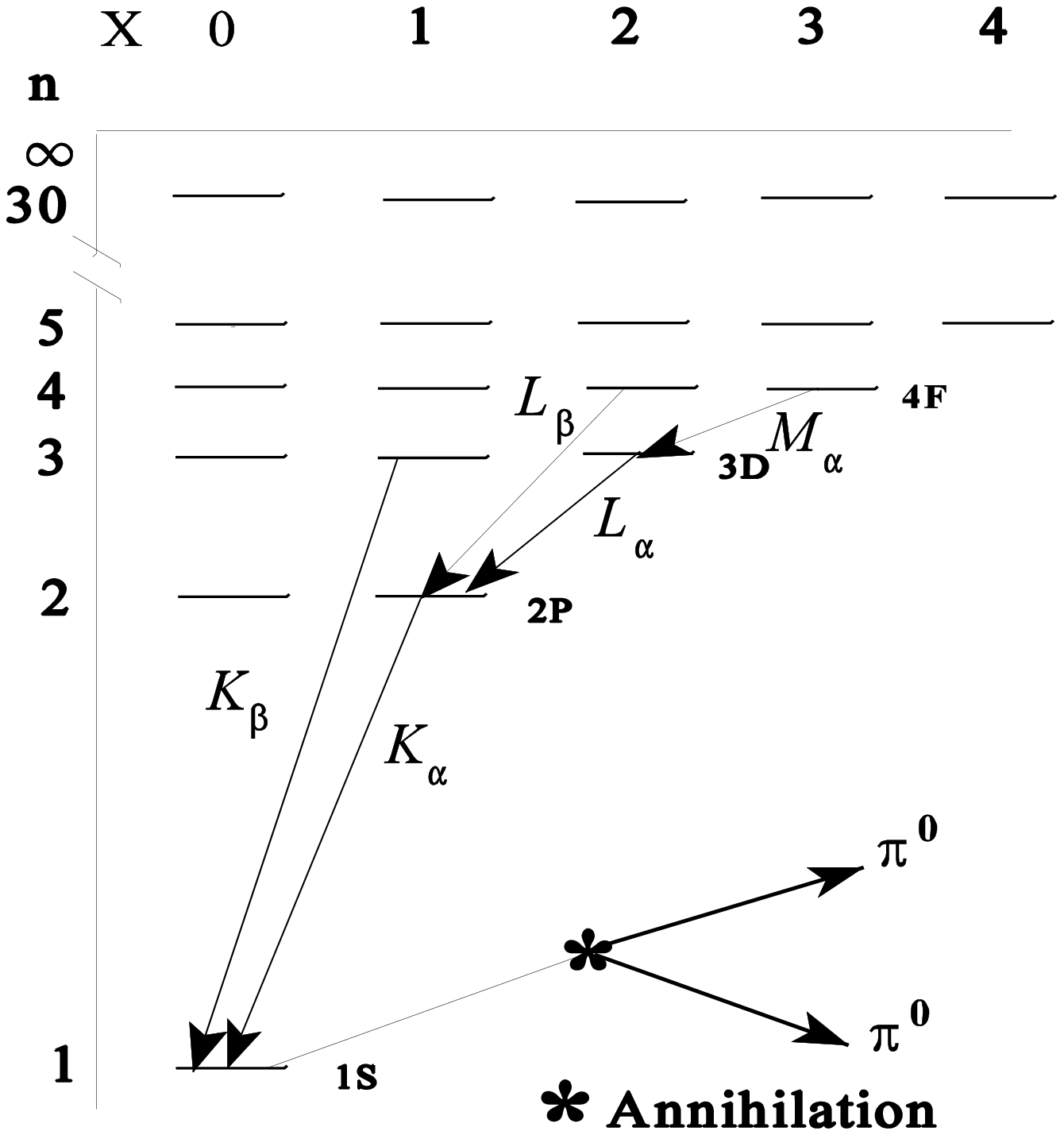}

\begin{figure}[t]
\caption{Levels of atomic orbital states for protonium with 
principle quantum number $n$ and orbital angular momentum $X$. 
It shows radiative cascades to the 1S state which 
involve K and L X-rays. Detection of triple coincidences 
of K and L X-rays 
and annihilation into $\pi^0 \pi^0 $ can prove that the reaction
occurs from an atomic S state.}
\label{f2}
\end{figure}

Assume the $\pi^0$ field 
is given by,

\begin{eqnarray}
\phi( x) = \sum_{\bf p} f(\omega_p) \left\{
\left[ \chi({\bf p}) - \eta({\bf p}) \right]e^{i {\bf p \cdot x}}
\right. \nonumber \\  \left.
+ \left[ \chi^\dagger({\bf p}) - \eta^\dagger({\bf p}) 
\right]e^{-i {\bf p \cdot x}} \right\},
\label{eqnphipi0}
\end{eqnarray}
where  $\chi^\dagger({\bf p})$ and $\eta^\dagger({\bf p})$ 
are creation
operators for two different $\pi^0$ states. 

If the $\pi^0$'s are in different states, there is no requirement 
that the state of two $\pi^0$'s must be symmetric under interchange.  
Therefore, $Y$ can be odd for some $\pi^0 \pi^0$ states
just as for the $\pi^+ \pi^-$ system, leading to a $P_1$ state with
parity of $-1$.

Next we will consider what is required for a state of two 
$\pi^0$'s to have charge parity of $-1$. Consider a particular
state of two $\pi^0$'s in the center of mass system,
\begin{equation}
\Phi = \sum_{\bf K} e^{i {\bf K \cdot R}}
\chi^\dagger({\bf K}) \eta^\dagger({\bf -K}) \Phi_0,
\label{eqnphi1}
\end{equation}
where  $\Phi_0$ is the vacuum state, ${\bf K}$ is the momentum of
each particle, and $ {\bf R = r_1 - r_2} $ is the relative
coordinate. Applying the charge parity operator, 
\begin{equation}
C\Phi = \sum_{\bf K} e^{i {\bf K \cdot R}}
C\chi^\dagger({\bf K})C^{-1} C \eta^\dagger({\bf -K})C^{-1}  \Phi_0.
\label{eqnphi1a1}
\end{equation}
 
If $\chi^\dagger({\bf p})$ and $\eta^\dagger({\bf p})$ were to each
change into themselves under $C$, then this state would be unchanged
and the eigenvalue of charge parity would still be $+1$. We need,
\begin{eqnarray}
C\chi^\dagger({\bf K})C^{-1} =  - \eta^\dagger({\bf K}), \nonumber \\ 
C\eta^\dagger({\bf K})C^{-1} = - \chi^\dagger({\bf K}),  
\label{eqncop}
\end{eqnarray}
for then,
\begin{equation}
C\Phi = \sum_{\bf K} e^{i {\bf K \cdot R}}
\eta^\dagger({\bf K}) \chi^\dagger({\bf -K})  \Phi_0.
\label{eqnphi2}
\end{equation}
But $\chi^\dagger({\bf p})$ and $\eta^\dagger({\bf p})$ commute. 
Thus, after letting $ {\bf K} \rightarrow -{\bf K}$ we obtain,
\begin{equation}
C\Phi = \sum_{\bf K} e^{-i {\bf K \cdot R}}
\chi^\dagger({\bf K}) \eta^\dagger({\bf -K})  \Phi_0.
\label{eqnphi3}
\end{equation}

Applying the coordinate exchange operator $P_r$ results in,
\begin{equation}
P_r C\Phi = \sum_{\bf K} e^{i {\bf K \cdot R}}
\chi^\dagger({\bf K}) \eta^\dagger({\bf -K})  \Phi_0 = \Phi.
\label{eqnphi4}
\end{equation}
Since $P_r^{-1} = P_r, C\Phi = P_r \Phi$. The eigenvalues of
$P_r = \pm 1$ depending upon the relative angular momentum of
the two particles. Thus, the charge parity becomes
the same as in the $\pi^+ \pi^-$ case, given by 
Eq. (\ref{eqnpipi}). If the $\pi^0 $ field
satisfies Eqs. (\ref{eqnphipi0}) and (\ref{eqncop}), 
then the annihilation reaction,
\begin{equation}
\overline p p \: (^3S_1) \rightarrow \pi^0 \pi^0 (P_1) 
\label{eqnpi0new} 
\end{equation}
is allowed. 

According to the Standard Model, 
the pion is composed of 
{\it up} and {\it down} quarks,

\begin{eqnarray}
\pi^+ = (u \: \overline d),&
\pi^- = (\overline u  \: d),&
\pi^0 = {1 \over \sqrt{2}}
(u \: \overline u - d \: \overline d ).
\label{eqnpiquark}
\end{eqnarray}
Although the $\pi^0 $ has two different states (i.e.,
$u \: \overline u$ and  $d \: \overline d$), the different
states do not change into each other under charge parity as required
by Eq. (\ref{eqncop}). Thus some change to the pion model is
needed. We cannot make any suggestions in that regard other 
than to note that composite states, such as
$g \: \overline h $ and $h \: \overline g$,
where $g$ and $h$ are two different fermions, satisfy 
Eq. (\ref{eqncop}). 

If the neutral pions are not identical just as the charged pions, 
then the branching ratio of 
$\overline{p} p \rightarrow  \pi^0\pi^0 $ from S states
should be half that of $\overline{p} p \rightarrow  \pi^+\pi^- $
by charge independence.
However, since the neutral pions have two different states they 
will only be non-identical half the time. Therefore, assuming all
annihilations are from S states, the branching ratio of
$\overline{p} p \rightarrow   \pi^0\pi^0 $ should equal 
$0.25 \times (31 \times 10^{-4})  = 7.8  \times 10^{-4}$ 
which is roughly 
in agreement with the Crystal Barrel collaboration's result of 
$(6.93 \pm 0.43) \times 10^{-4}$. 

A referee has pointed out that some of vector mesons 
could decay into two $\pi^0$'s if they are not identical,
providing another test of this theory. 
As discussed above two non-identical $\pi^0$'s 
can be in a $P_1$ state with $J = 1$, 
parity $= -1$, and charge parity $= -1$ which matches the
values for some of the vector mesons.
 
We considered the $\pi^0\pi^0$ decay mode of the $\rho(770)^0$,
$\omega(782), \phi(1020),$ and $J/\psi(1S)$. 
These reactions are forbidden by isospin 
conservation, but they can proceed electromagnetically.
In estimating the branching ratios, we assumed that the 
$\pi^0\pi^0$ mode would occur at about the same rate as the $\pi^+\pi^-$ 
mode, but reduced by a factor of two because the two $\pi^0$'s 
are only non-identical half the time. However, the $\rho(770)^0$
has $I = 1$, so the reaction $\rho(770)^0 \rightarrow \pi^+\pi^-$ 
is allowed by isospin conservation because 
the $\pi^+\pi^-$ system can have I $=$ 0, 1, or 2, while 
the reaction $\rho(770)^0 \rightarrow \pi^0\pi^0$ 
is forbidden because the $\pi^0\pi^0$ system can only have I $=$ 0 or 2. 
Occurring electromagnetically, its branching ratio would be reduced 
by the factor $\alpha^2$. 

Our estimated branching ratios are,
\begin{eqnarray}
BR(\rho(770)^0 \rightarrow  \pi^0\pi^0) = 5 \times 10^{-5}, \nonumber \\
BR(\omega(782) \rightarrow  \pi^0\pi^0) = 1 \times 10^{-2}, \nonumber \\
BR(\phi(1020) \rightarrow \pi^0\pi^0) = 4 \times 10^{-5}, \nonumber \\
BR(J/\psi(1S) \rightarrow  \pi^0\pi^0) = 7 \times 10^{-5}.
\label{branching}
\end{eqnarray}
The only measured 
upper limit~\cite{achasov},
$BR(\phi(1020) \rightarrow  \pi^0\pi^0) 
< 4 \times 10^{-5}$, is just in the expected range. 

Helpful discussions with Prof. J. E. Kiskis are 
gratefully acknowledged.

\end{document}